\documentclass[prl,showpacs]{revtex4}
\usepackage{dcolumn}
\usepackage{graphicx}
\usepackage{latexsym}

\begin{document}

\title{Integrable Hierarchies and Information Measures}

\author{Rajesh R. Parwani}
\email{parwani@nus.edu.sg}
\affiliation{Department of Physics and University Scholars Programe, \\
National University of Singapore, Kent Ridge, Singapore.\\}

\author{Oktay K. Pashaev}
\email{oktaypashaev@iyte.edu.tr}
\affiliation{Department of Mathematics
\\
Izmir Institute of Technology, Urla-Izmir, 35430 Turkey}

\begin{abstract}
In this paper we investigate integrable models from the perspective of information theory,  exhibiting various connections. We begin by showing
that compressible hydrodynamics for a one-dimesional isentropic fluid, with an appropriately motivated information theoretic extension, is described by a general nonlinear Schrodinger (NLS) equation. Depending on the choice
of the enthalpy function, one obtains the cubic NLS or other modified NLS equations that have
applications in various fields. Next, by considering the integrable hierarchy
associated with the NLS model, we propose higher order information measures which include the Fisher measure
as their first member. The lowest members of the hiearchy are shown to be included in the expansion of a regularized Kullback-Leibler measure while, on the other hand, a suitable combination of the NLS hierarchy leads to a Wootters type measure related to a NLS equation with a relativistic dispersion relation. Finally, through our approach, we are led to construct an integrable semi-relativistic NLS equation.

\end{abstract}

\maketitle

\section{Introduction}

Integrable equations are fascinating not just because of their soliton solutions and the connections they make among different areas of mathematics, but also because they do describe real physical systems in some limit. An example is the cubic nonlinear Schrodinger equation,
\begin{equation}
i\psi_t + \psi_{xx} + 2\kappa^2 |\psi|^2 \psi = 0 \label{cubic}
\label{NLS1}
\end{equation}
which is of relevance in quantum optics, condensed matter physics and other areas. The basic equation (\ref{cubic}) can be modified while still preserving integrability, for example by adding to the right-hand-side of (\ref{cubic}) a term proportional to
\begin{equation}
Q= s\frac{(\sqrt{\rho})_{xx}}{\sqrt{\rho}} \, ,
\end{equation}
where $\rho = |\psi|^2$. Through a change of variables one can actually absorb that extra term and regain the form (\ref{cubic}) at the expense of redefined parameters \cite{PashaevLeeMPLA}. However this is possible only if parameter $s < 1$, whereas if $s > 1$ one ends up with a reaction-diffusion equation \cite{PashaevLeeMPLA}. Such $Q$ augmented NLS equations have appeared in plasma physics \cite{LeePashaevRogersSchief},
where they desribe transmission of uni-axial waves in a cold collisionless plasma subject to a transverse magnetic field.

The reason for using the symbol $Q$ is because such a term, often referred to as a ``quantum potential", appeared first in alternate ways of writing the usual linear Scrodinger equation of quantum mechanics \cite{mad,deBroglie,Bohm}. Consider the one dimensional time-dependent Schrodinger equation (we set the mass $m=1$),
\begin{equation}
i\hbar \psi_t + \frac{\hbar^2}{2}\psi_{xx} - U(x)\psi = 0 \label{Schrodinger}
\end{equation}
Then substituting into this equation the Madelung representation of the wave function
\begin{equation}
\psi = \sqrt{\rho}e^{\frac{i}{\hbar}S}\label{Madelung}
\end{equation}
decomposes it into two real equations,
\begin{eqnarray}
{S_t} + {1 \over 2} (S_{x})^2 + U -{{\hbar}^2  \over 2 } \frac{(\sqrt{\rho})_{xx}}{\sqrt{\rho}} &=& 0 \, ,  \label{qhj} \\
{\rho_t}  +  \left( \rho S_x \right)_x  &=& 0 \, .  \label{cont}
\end{eqnarray}
The first equation may be viewed as a generalisation of the usual Hamilton-Jacobi equation by the term with explicit $\hbar$ dependence, the quantum potential, encoding the quantum aspects of the theory. The second equation is the
continuity equation expressing the conservation of probability.

Several attempts have been made to motivate the form of $Q$ and thus obtain a derivation of Schrodinger's equation from classical dynamics \cite{Nelson}. Here we adopt an information theoretic perspective similar to that used in statistical mechanics and which is usually refered to as the ``maximum entropy method" \cite{Jaynes}. The idea is that if one has a system that has to be described probabilistically then, lacking any information of the detailed microscopic dynamics, one should choose the probability distribution with minimum bias. This is achieved by maximising an appropriate measure of uncertainty (entropy), such as the Gibbs-Shannon measure used in classical statistical mechanics.

In order to proceed with an information theoretic interpretation of (\ref{qhj}, \ref{cont}), it is useful to approach those equations through  a variational principle \cite{reg}: one minimises the action
\begin{equation}
\Phi = \int  \rho \left[ {S}_t  + {1 \over 2} (S_x)^2   + U \right] dx dt   \ + {{\hbar}^2 \over 8}   I_F \, \label{varmulp}
\end{equation}
with respect to the field variables $\rho$ and $S$. The positive quantity
\begin{equation}
I_F  \equiv  \int dx dt \  \ \rho  \left({\rho_x \over \rho} \right)^2 \,  \label{fish}
\end{equation}
resembles the ``Fisher information" measure used in statistics \cite{Fisher, Kullback}. Since a broader probability distribution $\rho(x)$ represents a greater uncertainty in $x$, so $I_F$ may be thought of as an inverse uncertainty measure.

The equations (\ref{varmulp}, \ref{fish}) were used in Ref.\cite{reg} to interpret Schrodinger's equation as follows:  First one notes that without the term $I_F$, varying Eq.(\ref{varmulp}) gives rise to the Hamilton-Jacobi equation describing a classical {\it ensemble}. The probability, $\rho(x)$ appears in this context because one supposes that there is uncertainty in our knowledge of the initial position of the particle. One then adopts the principle of maximum uncertainty \cite{Jaynes} to constrain the probability distribution $p(x)$ characterising the ensemble: we would like to be as unbiased as possible in its choice, consistent with our lack of information. That constraint  is implemented in (\ref{varmulp}) by minimising $I_F$ when varying the classical action: $ {\hbar}^2 /8 $ is the Lagrange multiplier.

It remains to explain why $I_F$ is chosen as the information measure in the above quantum mechanical context as opposed to say the Gibbs-Shannon measure. In information theory and statistical mechanics the Gibbs-Shannon measure is the simplest possibility that satisfies certain axioms that are deemed necessary in those contexts \cite{Jaynes}. Similarly one can derive the Fisher measure as the relevant quantity that satisfies axioms relevant for classical ensemble dynamics and hence appropriate for use in deriving Schrodinger's equation \cite{par1}

In this paper we would like to apply the above information theoretic reasoning to motivate the NLS (\ref{cubic}) and its various extensions. In the next section we first review the derivation of the action for a classical compressible fluid in one dimension. Then in Sect.3 we use information theoretic arguments to modify the action and so arrive at a general nonlinear Schrodinger equation. In Sect.4 we employ an expansion of the enthalpy function to obtain specific examples of the nonlinear Schrodinger equation. In Section 5 we consider the hierarchy associated with the NLS equation and use that to define a hierarchy of higher-derivative information measures. Then in Section 6 we relate the information hiearchy to other information measures in the literature and in Section 7 we use the information measures to construct NLS equations with relativistic dispersion relations. Our conclusion is in Section 8 while in the appendix we discuss the relationship between the Madelung representation of quantum mechanics and the complexified Burgers equation.

\section{Compressible Fluid in One Dimension }
In this paper we focus on a specific physical model, hydrodynamics, to illustrate our approach, though we believe that much of it can be generalised to other contexts.

The Euler and continuity equations for a one-dimensional compressible fluid
are
\begin{equation}
v_t + v v_x + \frac{1}{\rho}P_x = 0 \, ,
\label{Euler1}
\end{equation}

\begin{equation}
\rho_t + (\rho v)_x = 0
\label{Continuity1}
\end{equation}
where the hydrodynamical variables corresponding to the density of fluid, velocity of fluid and pressure have been denoted by $\rho (x,t), v(x,t)$ and $P(x,t)$ respectively.
In addition one has the thermodynamic equation \cite{LavrentievShabat}
 \begin{equation}
\frac{d{\S}}{dt} = \frac{\partial {\S}}{\partial t} + v {\S}_x = 0
\label{entropy}
\end{equation}
expressing the conservation of entropy, ${\S}$, if one assumes the absence of the heat exchange between parts of the medium. To complete the description of the dynamics one also needs an equation of state
\begin{equation}
P = P({\S}, \rho)
\end{equation}
whose concrete form depends on the properties of the fluid. For example, an ideal gas has
\begin{equation}
P = e^{ {\S} / c_V} \rho^{\gamma} \, ,
\end{equation}
where $\gamma = c_P/c_V$ (the Poisson adiabate) is the ratio of specific heat capacities at constant pressure and volume respectively.

We will consider an {\it isentropic} fluid which has a spacetime
constant ${\S}$ so that (\ref{entropy}) is automatically
satisfied. For such barotropic processes the pressure becomes a
function of density only,
\begin{equation}
P = P (\rho) \, .
\end{equation}
 It is convenient to introduce the enthalpy function defined through the relation
\begin{equation}
\frac{\partial}{\partial x} E(\rho) = \frac{1}{\rho}\frac{\partial}{\partial x}P (\rho)
\label{entalpypressure}
\end{equation}
which implies, for $\rho_x \neq 0$,
\begin{equation}
E'(\rho) = \frac{1}{\rho}P'(\rho)
\end{equation}
or
\begin{equation}
E(\rho) = \int^\rho _{\rho_0}\frac{dP}{\rho} \, .
\end{equation}
Then the system of equations (\ref{Euler1}),(\ref{Continuity1})
becomes
\begin{equation}
v_t + v v_x + (E(\rho))_x = 0 \, ,
\label{Euler2}\end{equation}
 \begin{equation}
\rho_t + (\rho v)_x = 0 \, .
\label{Continuity2}\end{equation}
This system may be written in Lagrangian form by first introducing the velocity potential
\begin{equation}
v(x,t) = S_x(x,t) \,
\end{equation}
then integrating the first equation once and introducing the enthalpy potential
\begin{equation}
E(\rho) = \frac{dV (\rho)}{d \rho}
\end{equation}
to get
\begin{equation}
S_t + \frac{(S_x)^2}{2} +  \frac{dV (\rho)}{d \rho} = 0 \, ,
\label{HJ3}\end{equation}
 \begin{equation}
\rho_t + (\rho S_x)_x = 0 \, .
\label{Continuity3}\end{equation}
The action for this system is
\begin{equation}
A = \int \left(\rho S_t + \frac{\rho (S_x)^2}{2} + V(\rho)\right) dx dt \label{fluidact}
\end{equation}
and equations (\ref{HJ3}),(\ref{Continuity3}) appear by varying this functional with respect to
$\rho$ and $S$. The resemblance of (\ref{fluidact}) to the classical part of (\ref{varmulp}) will be the starting point for the extension in the next section.

We note, for later use below, that when the enthalpy vanishes, $E=0$, the fluid equations (\ref{HJ3}),(\ref{Continuity3}) are invariant under a scaling, $\rho \to \alpha \rho$. That is, with $E=0$, the magnitude of the density does not matter, only its variation. When $E$ is not zero, the equations to be derived later become generalised nonlinear Schrodinger equations, with a sensitivity to the magnitude of $\rho$.

\section{Information-Theoretical Extension of Compressible Fluid Dynamics}

The action (\ref{fluidact}) gives the classical equations of motion for the fluid. Since the density $\rho$ informs us about the likelihood of finding the microscopic fluid elements at a certain region of spacetime, it plays a role analogous to the probability density in quantum mechanics. Thus from the density we may form an information measure $I$ that quantifies our knowledge of the microstates and we may demand, as in the previous section, that the equations of motion follow from (\ref{fluidact}) but constrained such that our uncertainty (information) is maximised (minimised). This will lead to modified hydrodynamics equations that depend on the form of information measure chosen in the procedure.

Now, the density  $\rho(x)$ is positive definite and if it is uniform it tells us that the underlying particles of the fluid could be anywhere: we have no information (maximum uncertainty). If the density is peaked somewhere, we know that a fluid particle is more likely to be there, that is we have gained information. Thus we require that our scalar information functional $I[\rho]$ have the property that it is positive definite and $I \to 0$ as $\rho \to$ a constant.

We prefer local equations, and so we may write $I$ as an integral over a density function $J(\rho)$,
\begin{equation}
I  =  \int dx dt \ \rho  J(\rho) \label{iform}
\end{equation}
Next we assume the density to be slowly varying and so do a derivative expansion,
\begin{equation}
J(\rho) = J_0(\rho) + \rho^{\prime} J_1(\rho) +  \rho^{\prime \prime} J_{21}(\rho)  + (\rho^{\prime})^2 J_{22}(\rho) + \mbox{higher derivative terms} ……\, , \label{jexp}
\end{equation}
where $J_0(\rho), J_1(\rho),...$ do not contain any derivatives. We assume that when (\ref{jexp}) is used in (\ref{iform}) the integrals are convergent term by term.

We also impose the strong condition that the information measure, $I$,  does not break the invariance of the $E=0$ equations of motion (\ref{HJ3}) and (\ref{Continuity3}) under the scaling of $\rho \to \alpha \rho$.  That is, although $E$ will generally break that invariance, we demand that the terms in the modified equations of motion that come from $I$ do not do so: the information measure is chosen to be neutral to the magnitude of $\rho$ but measures only local variations. So here we see the first difference between the contributions of our $I$ and $E$ : $I$ is "unbiased" towards the size of $\rho$.

In order to achieve our goal, we need to demand that $J(\rho)$ in (\ref{iform}) is scale invariant (note we already factored out a $\rho$ in the integral form of $I$). In that way the equation of motion terms that come from varying $I$ will be scale invariant. This is satisfied if (\ref{jexp}) has the form
\begin{equation}
J(\rho) = a_0 + a_1 \times (\rho^{\prime} / \rho) +  a_{21} \times (\rho^{\prime \prime} / \rho)  + a_{22} \times (\rho^{\prime} / \rho)^2 + ……\, , \label{jexp2}
\end{equation}
where $a_{k}, a_{kl},..$ for $k,l =1,2,...$ are constants. Using this in the integral that defines $I$, Eq.(\ref{iform}), and dropping constants and total derivatives, only the $a_{22}$ term survives to leading order and it gives precisely the Fisher information measure!

Recall that we still need to demand positivity of our information measure: that fixes the lagrange multiplier to be positive if we are minimizing the information. Fortunately, the Fisher measure already satisfies the other required property, that it vanishes  as $\rho \to $ a constant.

Note that in the fluid problem we have in general
$\rho \rightarrow \rho_0$ as $|x| \rightarrow \infty$, so that
\begin{equation}
\int (\rho (x,t) - \rho_0) dx = 1 \label{bdd} \, .
\end{equation}
With the choice of the Fisher information measure
\begin{equation}
I_F = \int \frac{(\rho_x)^2}{\rho} dx = 4 \int (\sqrt{\rho})_x (\sqrt{\rho})_x dx
\label{Fisher}
\end{equation}
as motivated above, the extension of the boundary condition (\ref{bdd}) from the usual case in quantum mechanics
does not modify the convergence properties of $I$.

Thus we have variational functional
\begin{equation}
A + \frac{\lambda^2}{8} I_F
\end{equation}
where $\lambda$ is the Lagrange multiplier and the equations of motion that follow are
 \begin{equation}
S_t + \frac{(S_x)^2}{2} +  \frac{dV (\rho)}{d \rho} - \frac{\lambda^2}{2} \frac{(\sqrt{\rho})_{xx}}{\sqrt{\rho}}=0 \, ,
\label{QHJ4}\end{equation}
 \begin{equation}
\rho_t + (\rho S_x)_x = 0 \, .
\label{Continuity4}\end{equation}
These equations may be combined into one complex equation through the inverse Madelung transformation to give the following general Nonlinear Schrodinger equation,
\begin{equation}
i\lambda\psi_t + \frac{\lambda^2}{2}\psi_{xx} - E(|\psi|^2)\psi = 0 \, . \label{gnls}
\label{GNLS}\end{equation}

In summary, although both the enthalpy function, $E$, and the information functional, $I$, will contribute $\rho$ dependent terms to the equations of motion, their structure and origin is in general different. Using the lowest order information measure with the properties described above we get a generalized NLS equation (\ref{gnls}) that depends on the form of $E$.  Using more generalised information measures will give further extensions of the NLS equations.

We remark also that unlike the quantum mechanics case \cite{par1}, the deduction of the Fisher measure above did not use the separability condition: rather here we assumed a convergent derivative expansion of the information density. See also \cite{np} for similar arguments used in the relativistic case.

\section{Integrable Cases}

In this section we study the simplest form of the function $E$ in (\ref{gnls}) that will give rise to integrable systems.
If $E$ as function of $\rho = |\psi|^2$ is analytic then
\begin{equation}
E(|\psi|^2) = E_0 + E_1 |\psi|^2 + E_2 |\psi|^4 + ... + E_n |\psi|^{2n}+...
\end{equation}
This equation implies that the pressure according to (\ref{entalpypressure}) is also an analytic
function of the form
\begin{equation}
P(\rho) = E_0 + \frac{1}{2} E_1 \rho^2 + \frac{2}{3} E_2 \rho^3 + ... \frac{n}{n+1} E_n \rho^{n+1}
+...
\end{equation}
At the lowest order of nonlinearity we get the Nonlinear Schrodinger Equation (NLS) with cubic nonlinearity
\begin{equation}
i\lambda\psi_t + \frac{\lambda^2}{2}\psi_{xx} - (E_0 + E_1 |\psi|^2)\psi = 0
\label{NLS}\end{equation}
This model is integrable for both signs of $E_1$. For $E_1 > 0$ it is defocusing (e.g. repulsive Bose gas)
and nontrivial soliton solutions exist only with nontrivial boundary conditions, so in this case $E_0 \neq 0$.
For $E_1 < 0$ we have the focusing case (e.g. attractive Bose gas) for which soliton solutions exist for vanishing boundary conditions; so in this case we can put
$E_0 = 0$. Both cases have applications in nonlinear optics describing pulse propagating in nonlinear media.
Consider the second case again: by rescaling space and time variables
$t' = t/\lambda$, $x' = \sqrt{2}x/\lambda$ and the coupling constant $E_1 = - 2 \kappa^2$
we may rewrite it in the form (we now skip all upperscripts)
\begin{equation}
i\psi_t + \psi_{xx} + 2 \kappa^2 |\psi|^2 \psi = 0 \, .
\label{NLS1}\end{equation} The ``semiclassical" or dispersionless
limit of this equation was studied in \cite{Jin} in relation to
shock wave propagation in nonlinear optics. The wave form of this
semiclassical limit is a NLS equation perturbed by a quantum potential \cite{ANZIAM}
\begin{equation}
i\psi_t + \psi_{xx} + 2 \kappa^2 |\psi|^2 \psi = \frac{|\psi|_{xx}}{|\psi|}\psi \, .
\label{NLSQP}\end{equation}
Then we can conclude that inclusion of information characteristics in the form of the Fisher measure, produces NLS (\ref{NLS1})
from dispersionless NLS (\ref{NLSQP}) and corresponding solitons of the first one from the shock waves of the second one \cite{ANZIAM}.

\section{Integrable NLS Hierarchy and Higher-derivative Information Measures}

It is well-known that one can construct a hierarchy of
higher-order differential equations that are related to the cubic
NLS (\ref{NLS1}) and its complex conjugate, which are still
integrable \cite{AKNS},
\begin{equation} i\sigma_3  \left (\begin{array}{clcr}\psi \\
\bar\psi \end{array} \right)_{t_{N}}= \cal{R}^{N} \left (\begin{array}{clcr} \psi\\
\bar\psi \end{array} \right) \label{NLShierarchy}\end{equation}
where $t_N$, $N = 1, 2, 3, ...$ is an infinite time hierarchy.
Here ${\cal R}$ is the matrix integro-differential operator - the
recursion operator of the NLS hierarchy -
\begin{equation} {\cal{R}} = i\sigma_3\left(\begin{array}{cccr}\partial_x+ 2\kappa^2 \psi
\int^x \bar\psi & -2\kappa^2 \psi \int^x \psi
\\ & \\-2\kappa^2 \bar\psi \int^x
\bar\psi&\partial_x+2\kappa^2\bar\psi \int^x \psi\end{array}
\right) \, .
\end{equation}
For the first few members of the hierarchy N = 1,2,3,4 this gives
\begin{equation}
\psi_{t_1} = \psi_x \, ,
\end{equation}

\begin{equation}
i \psi_{t_2} + \psi_{xx} + 2\kappa^2|\psi|^2 \psi = 0 \, ,
\end{equation}

\begin{equation}
 \psi_{t_3} + \psi_{xxx} + 6\kappa^2|\psi|^2 \psi_x = 0 \, ,
\end{equation}

\begin{equation}
 i\psi_{t_4} = \psi_{xxxx} + 2\kappa^2\left(2 |\psi_x|^2 \psi + 4|\psi|^2 \psi_{xx}
 + \bar\psi_{xx}\psi^2 + 3 \bar\psi\psi^2_x
 \right) + 6 \kappa^4 |\psi|^4 \psi .
\end{equation}
In the linear approximation, when $\kappa = 0$, the recursion operator is just the momentum operator
\begin{equation}
{\cal{R}}_0 = i \sigma_3 \frac{\partial}{\partial x}
\end{equation}
and the NLS hierarchy (\ref{NLShierarchy}) becomes the linear Schrodinger hierarchy
\begin{equation}
i\psi_{t_n} = i^n \partial^n_x \psi \, .
\end{equation}
Written in the Madelung representation it produces the complex Burgers hierarchy
(see Appendix for $n=2$ case and \cite{PashaevGurkan} for arbitrary $n$).

Let us look more explicitly at the fourth order flow for which the Hamiltonian is
\begin{equation}
H = \int \left[\bar\psi_{xx}\psi_{xx} - 8 \kappa^2 \bar\psi_x\psi_x \bar\psi \psi -
\kappa^2(\bar\psi_x^2 \psi^2 + \bar\psi^2\psi_x^2) + 2 \kappa^4 |\psi|^6 \right] dx \, .
\end{equation}
In the Madelung representation
\begin{equation}
\psi = \sqrt{\rho}e^{iS} = e^{R + iS}
\end{equation}
this becomes
\begin{eqnarray}
H = \int [ \frac{\rho^2_{xx}}{4\rho} - \frac{\rho_{xx}\rho^2_x}{4\rho^2}
+ \frac{\rho^4_x}{16 \rho^3}
 + \\\rho S^2_{xx} + \rho S^4_x + 2 \rho_x S_x S_{xx}
- 2\kappa^2\left(\frac{5}{4}\rho^2_x + 3\rho^2 S^2_x\right) + 2
\kappa^4 \rho^3] dx\end{eqnarray} In fact for configurations with
$S =const$ we have only contributions from the first three terms
which can be combined into
\begin{equation}
H = \int (\sqrt{\rho})_{xx} (\sqrt{\rho})_{xx} dx \, .
\label{Fisher4}\end{equation}
This may be considered as a higher order analog of the Fisher information measure $I_F$ (\ref{Fisher}).

Generalizing, the above linearized Schrodinger hierarchy suggests, after the substitution $S = const$,
the even order information measure hierarchy
\begin{equation}
I_2 = \int (\sqrt{\rho})_x (\sqrt{\rho})_x dx \label{i2} \, ,
\end{equation}

\begin{equation}
I_4 = \int (\sqrt{\rho})_{xx} (\sqrt{\rho})_{xx} dx \label{i4} \, ,
\end{equation}
$$.................$$
\begin{equation}
I_{2n} = \int (\sqrt{\rho})_{x..x} (\sqrt{\rho})_{x...x} dx \, ,
\end{equation}
Here all odd members  vanish because their integrands are total
derivatives. 

We will use the above information hierarchy in the next section to
construct relativistic NLS equations and exhibit links between
different information measures known  in the literature.

Before leaving this section, we compute the above information measures for the one soliton solution of the NLS equation. The measure is
\begin{equation}
\rho_\nu (x) = \frac{\nu}{2 \cosh^2 \nu x}
\end{equation}
which satisfies
\begin{equation}
\int^\infty_{-\infty} \rho_\nu (x) dx = 1 
\end{equation}
and is characterized by a real parameter $\nu$ so that
\begin{equation}
\lim_{\nu \rightarrow \infty} \rho_\nu (x) = \delta(x) \, .
\end{equation}
Then information measures (\ref{i2}), (\ref{i4}),... for this
distribution apart from numerical constants are simply
\begin{equation}
I_{2} = \nu^{2},\,\, I_{4} = \nu^{4},\,\, ...,\,\, I_{2\nu} =
\nu^{2n},....
\end{equation}

\section{Relation to Kullback-Liebler and Wooters Measures}

The Gibbs-Shannon entropy
\begin{equation}
I_{GS} = -\int \rho(x) \ln \rho(x) \ dx \label{gs}
\end{equation}
may be genelaised to the Kullback-Leibler information \cite{Kullback}
\begin{equation}
I_{KL}(p,r) = -\int \rho(x) \ln {\rho(x) \over r(x)} \ dx \, ,
\label{kl}
\end{equation}
where $r(x)$ is a reference probability distribution. If one
chooses the reference  distribution to be the same as $\rho(x)$
but with infinitesimally shifted arguments, that is $r(x) = \rho(x
+ \Delta  x)$, then to lowest order,
\begin{eqnarray}
I_{KL} ( \rho(x), \rho(x + \Delta(x)) &=& {- (\Delta x)^2  \over
2} I_F (\rho(x)) + O(\Delta x)^3  \, ,\label{connection}
\end{eqnarray}
that is, the Fisher measure is recovered as the lowest order term in the expansion.

One may further generalise the Kullback-Liebler information by
introducing a parameter $0< \eta <1$, as used for example in \cite{par2},
\begin{equation}
 M \equiv \int \rho(x)\ln \frac{\rho(x)}{(1-\eta)\rho(x) + \eta \rho(x + \eta L)} dx \, .
\end{equation}
This form is nonsingular even if the density vanishes at any point.
For $L<<1$ we have the expansion
$$
M = L^2 \frac{\eta^4}{2}\int \frac{\rho^2_x}{\rho} dx - L^3 (\frac{\eta^6}{3}- \frac{\eta^5}{4})
\int \frac{\rho^3_x}{\rho^2} dx -
$$
$$- L^4 \left[\frac{\eta^6}{24}\int \frac{\rho^2_{xx}}{\rho} dx +
(\frac{\eta^7}{3} - \frac{\eta^6}{9} - \frac{\eta^8}{4})\int \frac{\rho^4_x}{\rho^3}\right] + O(L^5)
$$
where a number of surface terms have been dropped after integration by parts. Let us look at the symmetrised measure
$$
M(+L) + M(-L) = L^2 \eta^4\int \frac{\rho^2_x}{\rho} dx -
$$
$$- L^4 \left[\frac{\eta^6}{12}\int \frac{\rho^2_{xx}}{\rho} dx +
2(\frac{\eta^7}{3} - \frac{\eta^6}{9} - \frac{\eta^8}{4})\int
\frac{\rho^4_x}{\rho^3}\right] + O(L^6) \, ,
$$
where as before, the lowest order term, proportional to $L^2$, is the Fisher measure.
By choosing the parameter $\eta $ to satisfy
\begin{equation}
\eta^2 - \frac{4}{3}\eta + \frac{3}{8} = 0 \, ,
\end{equation}
or $\eta = (2 \pm  \sqrt{5/8})/3$ we can rewrite the next,
$O(L^4)$ term,  as the higher-derivative information measure given
by Eq.(\ref{Fisher4}).

Thus the first two members of the information hierachy $I_{2n}$ we
proposed in Sect.(5) are contained in the Kullback-Leibler
information.

\section{Relativistic NLS Equations}

Now let us consider other ways of combining and using
the information measures. Take a Hamiltonian of the form
\begin{equation}
H = c_2 I_2 + c_4 I_4 + ... + c_{2n} I_{2n} + ...\label{hform}
\end{equation}
where the constant coefficients $c_i$ depend on the context. For
example, for low momenta one may expand the relativistic
dispersion relation $E = \sqrt{m^2c^4 + p^2c^2}$ to get
\begin{equation}
E = mc^2 + \frac{p^2}{2m} - \frac{p^4}{8m^3 c^2}+...
\end{equation}
This may be used to construct a ``semi-relativistic" Schrodinger equation
as a formal power series
 \begin{eqnarray}
 i\hbar \frac{\partial}{\partial t}\psi &=&
 mc^2(1 - \frac{\hbar^2}{2m^2c^2}\frac{\partial^2}{\partial x^2} - \frac{\hbar^4}{8m^4c^4}
 \frac{\partial^4}{\partial x^4}+ ...)\psi \\
 &\equiv & \hat{h} \psi
 \end{eqnarray}
 where
apart from a constant, the average of $\hat{h}$ for real $\psi$ is precisely
(\ref{hform}) for a particular choice of coefficients. In fact one
may proceed further: by replacing the derivative operator
$\partial/\partial x$ with the recursion operator $\cal{R}$ one
obtains an {\it integrable} relativistic nonlinear Schrodinger
equation

\begin{equation} i\sigma_3  \left (\begin{array}{clcr}\psi \\
\bar\psi \end{array} \right)_{t}= mc^2 \sqrt{ 1 + \frac{1}{m^2 c^2}{\cal{R}}^{2}} \left (\begin{array}{clcr} \psi\\
\bar\psi \end{array} \right) \label{RelNLS}\end{equation}

We note that relativistic versions of the Schrodinger equation
have been considered in different contexts, for example to study
relativistic quarks in nuclei \cite{Salpeter} and gravitational
collapse of a boson star \cite{frohlich}. A nonlinear version has
appeared as the semi-relativistic Hartree-Fock equation
\cite{Hartree}. But none of those models is known to be
integrable. By contrast the model (\ref{RelNLS}), where the
square root is considered as a formal power series (matrix
pseudo-differential operator), is an integrable nonlinear Schrodinger
equation with relativistic dispersion:
\begin{equation}
i \psi_t = m c^2 \sqrt{1 - \frac{1}{m^2
c^2}\frac{\partial^2}{\partial x^2}}\, \psi + F(\psi)
\end{equation}
where the nonlinearity expanded in $1/c^2$ is the infinite sum
\begin{eqnarray}
\lefteqn{F(\psi) = \frac{1}{2m}[-2\kappa^2 |\psi|^2 \psi]} \nonumber
\\
& & - \frac{1}{8m^3
c^2} [2\kappa^2(2|\psi_x|^2\psi + 4|\psi|^2\psi_{xx} +
\bar\psi_{xx}\psi^2 + 3\bar\psi \psi^2_x) + 6\kappa^4 |\psi|^4
\psi]  \nonumber \\
& & + O(\frac{1}{c^4}) \, .
\end{eqnarray}

What is amazing is that if we expand also the dispersion part in $1/c^2$,
then at every order of $1/c^2$  we get an integrable system. It
means that  we have integrable relativistic corrections to the NLS
equation at any order.

 Another way of constructing a relativistic model that
includes higher-derivative information measures is to  use
rapidity variables for the relativistic dispersion relation,
\begin{equation}
E = mc^2 \cosh \chi,\,\,\,p = mc \sinh \chi \ ,
\end{equation}
This gives the relativistic model with Hamiltonian \cite{mirkr}
\begin{equation}
H = mc^2 \int \bar\psi \cosh (L\frac{\partial}{\partial x}) \,
\psi dx\end{equation} Expanding the $\cosh$ we again have a member
of the information hierarchy and a relativistic NLS equation.

Finally, the above free Hamiltonian may be represented as a finite
difference operator
\begin{equation}
H = mc^2/2 \int \bar\psi (e^{L\frac{\partial}{\partial x}}+
e^{-L\frac{\partial}{\partial x}}) \psi dx = \int (\bar\psi(x)
\psi(x+L) + \bar\psi (x)\psi(x-L)) dx \, .
\end{equation}
The dispersive part of this hierarchy for $S = const$ gives a Wootters type \cite{Wooters} measure
\begin{equation}
I_W = \int (\sqrt{\rho(x)}\sqrt{\rho(x+L)} +  \sqrt{\rho(x)}\sqrt{\rho(x-L)})dx \, .
\end{equation}

\section{Summary}
We have shown how information theory arguments can be used to motivate the general nonlinear Schrodinger equation in the context of hydrodynamics. This then led us to study different information measures.

We noted that the integrable hierarchy of
linear and nonlinear Schrodinger equations, in their Madelung
form, naturally suggest a hierarchy of information measures of which the
Fisher measure represents the first member.
 The lowest members of the information hiearchy were shown to be included in
the expansion of a regularized Kullback-Leibler measure.

We also showed that how to contruct integrable semi-relativistic nonlinear Schrodinger equations using various combinations of the information measures. These classes of equations, which are distinct from those obtained in \cite{np} and references therein,  might be useful in analyzing relativistic
corrections to solitons, Bose-Einstein condensates or other condensed matter systems with effective equations of relativistic form.\\

{\bf Acknowledgment} This work was supported in part by the National
University of Singapore and the Izmir Institute of Technology, Turkey.

\section*{Appendix A: Madelung Fluid and Complex Burgers Equation}

Here we show that the Madelung representation used in quantum
mechanics may be viewed as a complexified version of the Cole-Hopf
transformation \cite{cole}, \cite{hopf} which relates the linear
heat equation with the nonlinear Burgers equation \cite{burgers}.

Begin with the  one dimensional time-dependent Schrodinger equation,
\begin{equation}
i\hbar \psi_t + \frac{\hbar^2}{2m}\psi_{xx} - U(x)\psi = 0 \label{Schrodinger}
\end{equation}
and define complex velocity
\begin{equation}
u(x,t) = u_c + i u_q = \frac{\hbar}{i m}(\ln \psi)_x \label{CCH}
\end{equation}
we have
\begin{equation}
u_c = \frac{1}{m}S_x,\,\,\,u_q = -\frac{\hbar}{2m} (\ln \rho)_x \label{velocity}
\end{equation}
This show that the real part of the complex velocity is the
classical velocity, while the imaginary part can be associated
with the "quantum velocity" \cite{salesi}. Considering (\ref{CCH})
as complex analog of the Cole-Hopf relation we can see that the
Schrodinger equation (\ref{Schrodinger}) is equivalent to the
complex Burgers equation
\begin{equation}
u_t + u u_x = \frac{i\hbar}{2m}u_{xx} - \frac{1}{m} U\label{Burgers}
\end{equation}
Splitting this equation into real and imaginary parts, for the former
we have a hydrodynamic equation for the classical velocity
\begin{equation}
{u_c}_t + u_c {u_c}_x = - \frac{1}{m}(U + \frac{\hbar}{2} {u_q}_x - \frac{m}{2}{u_q}^2)_x\label{Madelung1}
\end{equation}
while for the imaginary part we obtain the transport equation for the
quantum velocity
\begin{equation}
{u_q}_t + (u_c u_q)_x = \frac{\hbar}{2m}{u_c}_x\label{Madelung transport}
\end{equation}
After one space integration the last equaton becomes the continuity equation
\begin{equation}
\rho_t + (\rho u_c)_x = 0\label{Madelung2} \, .
\end{equation}
Equations (\ref{Madelung1}),(\ref{Madelung2}) describe a Madelung
fluid. Thus the above consideration show that a Madelung fluid can
be described by one nonlinear complex Burgers equation
(\ref{Burgers}). In equation (\ref{Madelung1}) the potential part
defines a quantum potential written through the Riccati equation
\begin{equation}
U_q = \frac{\hbar}{2} {u_q}_x - \frac{m}{2}{u_q}^2\label{QP} \, .
\end{equation}
Then expression (\ref{velocity}) for the quantum velocity
\begin{equation}
u_q = -\frac{\hbar}{2m}\frac{\rho_x}{\rho}
\end{equation}
gives a linearization of the Riccati equation
\begin{equation}
(\sqrt{\rho})_{xx} + \frac{2m}{\hbar^2} U_q (\sqrt{\rho}) = 0\label{linearRiccati}
\end{equation}
with potential
\begin{equation}U_q = -\frac{\hbar^2}{2m}\frac{(\sqrt{\rho})_{xx}}{\sqrt{\rho}} = -\frac{\hbar^2}{2m}\frac{|\psi|_{xx}}{|\psi|} \label{QP2} \, .
\end{equation}

Finally we note that equation (\ref{linearRiccati})
has simple geometrical interpretation. If
\begin{equation}
ds^2 = dt^2 + |\psi|^2 dx^2\label{metric}
\end{equation}
is the metric of a surface with geodesic coordinates (t,x), then the Gaussian curvature
becomes the quantum potential
\begin{equation}
K = - \frac{(\sqrt{\rho})_{xx}}{\sqrt{\rho}} = \frac{2m}{\hbar^2} U_q\label{GC}
\end{equation}
This shows that the quantum potential can be interpreted as the Gaussian curvature
\begin{equation}
U_q = \frac{\hbar^2}{2m} K \, .
\end{equation}
In particularly it implies that the classical limit corresponding
to a vanishing quantum potential locally is  a plane.

\end{document}